# Radial modal transitions of Laguerre-Gauss modes during parametric upconversion: towards the full-field selection rule of spatial modes

Hai-Jun Wu (吴海俊),[1] Li-Wei Mao (毛立伟),[1] Yuan-Jie Yang (杨元杰),[3] Carmelo Rosales-Guzmán,[1] Wei Gao (高玮),[1] Bao-Sen Shi (史保森),[1,2] and Zhi-Han Zhu (朱智涵)[1,*]

[1] *Wang Da-Heng Collaborative Innovation Center, Heilongjiang Provincial Key Laboratory of Quantum Manipulation & Control, Harbin University of Science and Technology, Harbin 150080, China*
[2] *CAS Key Laboratory of Quantum Information, University of Science and Technology of China, Hefei, 230026, China*
[3] *School of Physics, University of Electronic Science and Technology of China, Chengdu, China*

Optical orbital angular momentum transformation and corresponding azimuthal-mode selection rules have been studied exhaustively for various nonlinear optical interactions. However, nonlinear transformation of radial mode has not been systematically studied since the pioneering work [Phys. Rev. A 56, 4193, 1997]. In this paper, we theoretically study and experimentally verify the radial modal transitions of Laguerre-Gauss (LG) modes in parametric upconversion. Specifically, we provide a general solution that describes the sum-frequency generation (SFG) field excited by two arbitrary LG modes. Based on the solution, one can predict the full spatial complex amplitude of SFG fields upon propagation precisely and readily obtain the associated full-field selection rule including both azimuthal and radial modes. This work provides a theoretical basis for quantum and nonlinear optical research involving parametric upconversion of complex structured light, and paves the way for future work on full-field transformation of spatial modes in other nonlinear interactions.

*Introduction.* — Soon after Allen *et al.* discovered optical orbital angular momentum (OAM) [1], research on the transformation of OAM in nonlinear optics began with the second-harmonic generation (SHG) of Laguerre-Gaussian (LG) modes reported by Dholakia *et al.* in 1996 [2]. The azimuthal modal transition identified in their work, i.e., that SHG fields carry twice the azimuthal indices of the pump, provided straightforward insight into OAM conservation during nonlinear interactions at the photon level. One year later, the researchers further reported an SHG excited by LG modes with radial indices [3], in which they found that the radial structure of the SHG fields varied upon propagation, and it was more interesting that the beam profiles at the generation plane and the far field were same. Thereafter, perhaps because OAM conservation is of interest for both practical applications and fundamental optics, only the transformation of azimuthal indices and corresponding OAM selection rule was studied comprehensively and thoroughly in various nonlinear optical processes. Research included second-order interactions, third-order interactions, high order harmonic generation, and even light-sound (or other matter wave) interactions [4–12]. On the basis of these OAM selection rules, in quantum domain, nonlinear interactions involving photonic OAM have become crucial to high-dimension entanglement generation, memory, and frequency conversion [13–15].

Recently, to exploit spatial degrees of freedom (DoFs) fully, research on structured light has gradually begun to focus on the forgotten radial mode [16–24], particularly for their potential value in quantum information [25, 26]. With regard to this, generation, memory, and frequency conversion of light with on-demand full transverse structure are important. The premise for these tasks is to know the *full-field* selection rule of spatial modes that control the transformation of transverse structure in nonlinear interactions, e.g., for LG modes is to know both azimuthal and radial modal transitions. However, after the pioneering work reported by Courtial *et al.* [3], research on radial modal transitions in nonlinear optics has rarely been considered. More recently, although a few studies have reported relevant interesting phenomena, such as the dependence between the azimuthal modal and the radial modal in parametric processes [27–31], there is no unambiguous *full-field* selection rule of spatial modes for a specific nonlinear process thus far.

Here, driven by the above research requirement and interest, we revisit the radial modal transitions of LG modes during parametric upconversion, or rather sum-frequency generation (SFG). On the theoretical side, we provide a general solution of the SFG field pumped by two arbitrary LG modes, which can precisely describe the spatial complex amplitude (including both azimuthal and radial structures) of the SFG field from the generation plane to the far field. Based on this solution, the *full-field* selection rule can be readily obtained via well-established state tomography based on projective measurement. On the experimental side, we verify the theory with the help of complex amplitude modulation and digital propagation [32–34]. All simulations are successfully observed via propagation tomography.

*Theory.* — The traveling wavefunction of a LG mode in cylindrical coordinates can be fully determined by knowing the wave vector $k$, beam waist $w_0$ at the original plane, and two spatial (i.e., azimuthal and radial) indices $\ell$ and $p$ given by [1]:

[*] zhuzhihan@hrbust.edu.cn

$$LG_p^\ell(r,\varphi,z) = g(r,z;\ k,w_0,p,\ell)L_p^{|\ell|}(\gamma)e^{-i\ell\varphi}, \qquad (1)$$

where $g(\cdot)$ denotes a Gaussian envelope function, $e^{-i\ell\varphi}$ is the twisted phase that gives rise to the photonic OAM, $L_p^{|\ell|}(\gamma)$ is the Laguerre polynomial with mode orders $p$ and $|\ell|$, and the variable $\gamma = 2r^2/w_0^2[1+(z/z_R)^2]$ ($z_R = kw_0^2/2$ is the Rayleigh length). It should be noted that $L_p^{|\ell|}(\gamma)$ provides the radial structure of the spatial amplitude. To be more specific, the zeros of $L_p^{|\ell|}(\gamma)$ specify the number ($N = p \geq 0$) and position of phase dislocations along the radial coordinate, thus leading to a specific multi-ring structure.

Of the two spatial indices, the azimuthal index $\ell$ (also known as the topological charge) can be regarded as the eigenvalue of the OAM operator with respect to the $z$-axis $\hat{L}_z = -i\hbar(\partial/\partial\varphi)$, i.e., $\hat{L}_z LG_p^\ell(r,\varphi,z) = \ell LG_p^\ell(r,\varphi,z)$ because of the twisted phase $e^{-i\ell\varphi}$ within the LG modes. Here, the operator $\hat{L}_z$ depends only on the DoF of $\varphi$, such that $\ell$ is a conserved quantity in paraxial propagation and the corresponding OAM selection rule in rotational-symmetric nonlinear interactions can be readily inferred. The operator of the radial index $p$ can be defined via a relation $\hat{P}_z LG_p^\ell(r,\varphi,z) = p LG_p^\ell(r,\varphi,z)$, however, this operator, unlike $\hat{L}_z$, is correlated with other observable quantities including $w_0$, $\ell$, and even $z$ [18]. That is, the radial quantum number can only be well defined when all of the relevant beam parameters are known. This is why controlling the radial mode experimentally is more difficult than controlling the azimuthal mode. Thus, most recent experiments that explore the radial quantum number have used imaging systems to match the relevant beam parameters between the sender and receiver [19–21]. Precisely because of this complexity, except when interacting with a plane wave, revealing the *full-field* selection rule that determines both azimuthal and radial modal transitions of LG modes during nonlinear interactions remains a challenge. Here, we start from parametric upconversion to investigate the *full-field* transformation of LG modes in general SFG.

The full theoretical framework for a general SFG pumped by two arbitrary LG modes is provided in Appendix A. In the main text, for simplicity and without loss of generality, we assume that the SFG is pumped by two beams with the same $k$ and $w_0$ but orthogonal polarizations in a type-II crystal, i.e., a special SFG known as type-II SHG. If we further assume the SHG is pumped by two collimated and collinearly propagating LG modes $LG_{p_1}^{\ell_1}$ and $LG_{p_2}^{\ell_2}$ in the perfect phase-matching condition. As the source of the SHG field, the nonlinear polarization (NP) excited by them at the generation plane ($z_0$) can be expressed as $\mathbf{P}^{2\omega} = \kappa E^{2\omega}(r_0,\varphi_0)$, where $\kappa = \epsilon_0 \chi^{(2)}$ denotes the nonlinear coupling coefficient and $E^{2\omega}(r_0,\varphi_0) = LG_{p_1}^{\ell_1}(r_0,\varphi_0)LG_{p_2}^{\ell_2}(r_0,\varphi_0)$ is the quadratic beating field. From this NP, one can readily infer that the SHG undergoes a doubling transition in the longitudinal-mode DoF (or frequency), i.e., $\exp[ik(\omega)z] \to \exp[ik(2\omega)z]$.

Moreover, this NP also fully determines the transition in the transverse-mode DoF (or spatial mode) involving selection rules for both azimuthal and radial modes. Note that, for the broadband phase matching of SHG, the subtle effect of spatial dispersion of high-order LG modes on the phase matching can be neglected [35, 36]. For the azimuthal mode, if the two pumps all carrying single-valued topological charge, one can obtain a concise azimuthal modal transition, i.e., $\exp(i\ell_1\varphi)\exp(i\ell_2\varphi) \to \exp[i(\ell_1+\ell_2)\varphi]$, and associated OAM selection, i.e., $\ell_{SHG} = \ell_1 + \ell_2$. For the radial mode, at this stage we can conclude only that the radial structure of the amplitude at $z_0$ plane is governed by the term $L^{2\omega}(r,z_0) = L_{p_1}^{|\ell_1|}(\cdot)L_{p_2}^{|\ell_2|}(\cdot)$ (see Eq. S2 in Appendix A). Except when $p_1$ and $p_2$ are zero simultaneously, the generated SHG is not an eigen solution of the paraxial wave equation, leading to an unstable radial amplitude upon propagation. Therefore, to reveal the complete radial amplitude of the SHG upon propagation, it is necessary to obtain the traveling wavefunction $E^{2\omega}(r,\varphi,z)$.

The general solution of $E^{2\omega}(r,\varphi,z)$ can be derived using the Collins propagator with $E^{2\omega}(r_0,\varphi_0)$ as the pupil function [37, 38], given by [see Appendix A for details]:

$$\begin{aligned} E^{2\omega}(r,\varphi,z) &= g^{2\omega}(r,z;\ 2k,w_0,p_{1,2},\ell_{1,2})L^{2\omega}(r,z)e^{-im\varphi} \\ L^{2\omega}(r,z) &= \sum_{j=0}^{p_1+p_2} q_j(z)L_{n+j}^{|m|}(\zeta_z) \end{aligned} \qquad (2)$$

where $g^{2\omega}(\cdot)$ is the amplitude envelope, two parameters $m = \ell_1 + \ell_2$, and $n = (|\ell_1|+|\ell_2|-|\ell_1+\ell_2|)/2$, and $L^{2\omega}(r,z)$ is a series function about $z$ that governs the radial structure of the SHG field upon propagation. The term $e^{-im\varphi}$ contained in Eq. (2) indicates that $E^{2\omega}(r,\varphi,z)$ carries a well-defined OAM of $m = \ell_1 + \ell_2$ per photon that corresponds to a concise OAM selection rule. For the radial mode, we note that the governing term $L^{2\omega}(r,z)$ consists of a series of Laguerre polynomials $q_j(z)L_{n+j}^{|m|}(\zeta_z)$. In addition, the weighting coefficient $q_j(z)$ and the complex variable $\zeta_z$ are both functions of $z$. This indicates that the radial amplitude structure of the SHG field is *usually* (except for $p_1 = p_2 = 0$) not constant upon propagation. Using Eq. (2), one can exactly simulate the full spatial complex amplitude of the SHG field from $z_0$ plane to the far field ($z_\infty$). Furthermore, by reformulating Eq. (2) as a coherent superposition of eigen LG modes, we can obtain the *full-field* selection rule of LG modes involving both azimuthal and radial modes, given by,

$$E^{2\omega}(r,\varphi,z) = \sum_{\eta=0}^{n+j} a_\eta LG_\eta^m(r,\varphi,z), \qquad (3)$$

where $a_\eta$ denotes weight coefficients that can be obtained by projecting Eq. (2) onto the corresponding eigenmode. Moreover, $a_\eta$ are real number because $\zeta_z$ in Eq. (2) is also real number at $z_0$ plane, which indicates the initial intramodal phase can only be $\pm\pi$. Eq. (3) represents a radial modal degenerate vortex beam (similar to Hypergeometric-Gaussian beams [39, 40]), and note that the mode orders of each components, i.e., $2\eta + m$, are different from each other. This leads to different speed in accumulation of Gouy phase upon propagation. That is to say, Eq. (3) explains, via another perspective, why the radial amplitude structure of the SHG field is *usually* (except for the special case of only one $LG_\eta^m$ term) propagation variant. Additionally, from the polynomial order $n+j$ in Eq. (2) and $\eta$ in Eq. (3), we know that the radial-mode components of the SHG field depends on both the azimuthal and radial indices of two pump fields.

From Eqs. (2) and (3) we see that, except for $z_0$ plane, the radial governing term $L^{2\omega}(r,z)$ and the coefficients $a_\eta$ will become real again at the far field due to $\zeta_z \to \mathbb{R}$ at $z_\infty$. At the far field, Eq. (2) becomes a Fourier transformation of $E^{2\omega}(r_0, \varphi_0)$, given by

$$E^{2\omega}(r,\varphi,z_\infty) = \mathcal{F}[E^{2\omega}(r_0,\varphi_0)] \\ = g^{2\omega}(r,z_\infty; w_0, p_{1,2}, \ell_{1,2}) L^{2\omega}(r,z_\infty) e^{-im\varphi} \quad (4)$$

Here, $L^{2\omega}(r,z_\infty)$ is a one-dimensional series function of $r$ that governs the far-field radial structure. Remarkably, for the particular case $p_1 = p_2 = p$, it can be further factorized into two concise forms:

$$\begin{cases} L^{2\omega}(r,z_\infty) = L_p^{|\ell_1|}(\bullet) L_p^{|\ell_2|}(\bullet), & \text{for } \ell_1\ell_2 > 0 \\ L^{2\omega}(r,z_\infty) = L_{p+n}^{|\ell_1+\ell_2|}(\bullet) L_p^{0}(\bullet), & \text{for } \ell_1\ell_2 < 0 \end{cases} \quad (5)$$

Equation (5) implies two interesting phenomena about the far-field amplitude: (i) if $\ell_1$ and $\ell_2$ have the same sign, the radial profiles at the $z_\infty$ and $z_0$ planes, except for overall enlarging upon propagation, are exactly same; and (ii) if $\ell_1$ and $\ell_2$ have opposite signs, the number of phase dislocations ($N$) at the far field is $N = 2p + n$. This phenomenon is associated with spatial cross-correlation of the two pumps. We further demonstrate the above predictions by analyzing simulations of the two cases.

For case (i), we first revisit the special case reported in Ref. 3, i.e., $\ell_1 = \ell_2 = \ell$ corresponding to the SHG pumped by a single LG mode. According to Eq. (5), the relation $L^{2\omega}(r/w_z, z_\infty) = L^{2\omega}(r/w_z, z_0) = [L_p^{|\ell|}(\bullet)]^2$ indicates that the radial profiles of the SHG field at $z_0$ and $z_\infty$ planes are exactly the same, i.e., $E^{2\omega}(r/w_z, z_0) = E^{2\omega}(r/w_z, z_\infty)$ and without any radial phase dislocations ($N = 0$). To illustrate this special case, Fig. 1(a) shows the simulated radial profile of the SHG pumped by $LG_2^0$, whose radial modal transition, according to Eq. (3), is $LG_2^0 LG_2^0 \to \sqrt{9/22} LG_0^0 + \sqrt{4/22} LG_2^0 + \sqrt{9/22} LG_4^0$. We then consider a slightly more general case of $|\ell_1| \neq |\ell_2|$, where we still produce the conclusion that $E^{2\omega}(r/w_z, z_0) = E^{2\omega}(r/w_z, z_\infty)$ but the number of phase dislocations is $N = 2p$. With regard to this prediction, Fig. 1(b) shows the simulated SHG pumped by $LG_2^1$ and $LG_2^2$ that corresponds to a radial modal transition of $LG_2^1 LG_2^2 \to \sqrt{1/6} LG_0^3 + \sqrt{5/27} LG_2^3 + \sqrt{35/54} LG_4^3$. Here, both in Figs. 1(a) and (b), the propagation distance is given in Gouy phase $\phi_g \in [0, 2\pi]$ so that one Rayleigh distance ($z_R$) is in the middle position. And we see that the radial profiles not only at the near- and far-fields are exactly the same, but also are mirror symmetry with respect to the $\phi_g = \pi/4$ (i.e., $z = z_R$) plane. This self-imaging propagation [41] is attributed to the SHG fields are radial degenerate superposition of LG modes with $p = 0, 2, 4$, and thus the evolution of intramodal phase structure is mirror symmetry with respect to $\phi_g = \pi/4$.

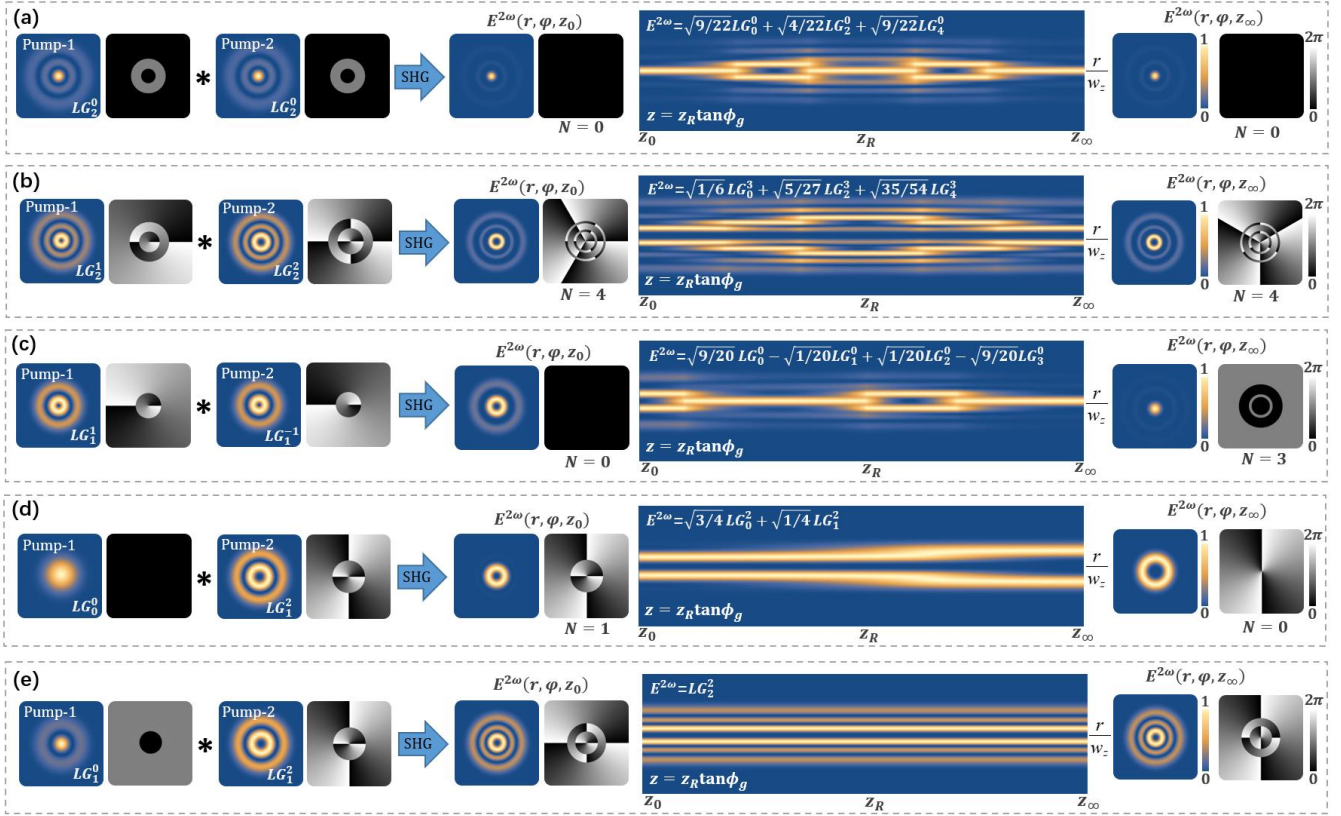

FIG. 1. Simulated spatial complex amplitudes of SHG fields of (a) $LG_2^0 LG_2^0$, (b) $LG_2^1 LG_2^2$, (c) $LG_1^1 LG_1^{-1}$, (d) $LG_0^0 LG_1^2$, and (e) $LG_1^0 LG_1^2$. In the radial profiles, the transverse unit is $r/w_z$ so that the beam enlarging upon propagation is compensated; and the propagation distance is given in Gouy phase $\phi_g \in [0, 2\pi]$, i.e., $z = z_R \tan\phi_g$, and thus one Rayleigh distance ($z_R$) is in the middle position.

For case (ii), we consider a special case of $\ell_1 = -\ell_2$, i.e., the case where the SHG is pumped by a pair of conjugated LG modes. Figure 1(c) shows the simulated SHG pumped by $LG_1^1$ and $LG_1^{-1}$, corresponding to a radial modal transition: $LG_1^1 LG_1^{-1} \to \sqrt{9/20}LG_0^0 - \sqrt{1/20}LG_1^0 + \sqrt{1/20}LG_2^0 - \sqrt{9/20}LG_3^0$. We see that the radial profiles are not mirror symmetry about $z = z_R$ and the number of radial phase dislocations at $z_\infty$ has an interesting value of $N = 2p + |\ell|$, which is exactly the mode order of $LG_p^{\pm\ell}$ [42]. The origin of this coincidence is the Fourier relation $\mathcal{F}[E^{2\omega}(r_0,\varphi_0)] = \mathcal{F}[LG_p^\ell(r_0,\varphi_0)LG_p^{-\ell}(r_0,\varphi_0)]$. We can thus regard the far-field amplitude of case (ii) as the auto (or cross) correlation of LG modes [43]. Furthermore, to put this in perspective of the radial-mode components, unlike case (i), here the highest-order $\eta = n + j$ is controlled by both azimuthal and radial indices of pump fields.

Beyond the particular cases obeying Eq. (5) discussed above, we seek to address a practical problem: whether one can use an easily-obtained Gauss beam to upconvert a 'signal' field encoded with the *full-field* spatial structure. Based on Eqs. (2) and (3), Fig. 1(d) shows the simulated SHG of $LG_0^0 LG_1^2$. The spatial amplitude of the SHG does not keep constant upon propagation because of a radial degenerate mode generation, i.e., $LG_0^0 LG_1^2 \to \sqrt{3/4}LG_0^2 + 1/2 LG_1^2$. In this case, the only feasible way to accomplish this task is using a near plane wave pump such as a flat-top beam. In addition, it is important to note that the propagation-variant SHG, corresponding to a radial degenerate superposed LG mode, demonstrated above is not always present. For instance, for the SHG pumped by $LG_1^2$ and $LG_1^0$ shown in Fig. 1(e), we see that the spatial amplitude keeps constant upon propagation, and this elegant result corresponds a concise *full-field* transformation: $LG_1^2 LG_1^0 \to LG_2^2$. For this type concise transformation, we can obtain the selection rule by solving $a_\eta = 1$ in Eq. (3), and for the special case $p_1 = p_2 = 1$ is given by, $LG_1^{(n-1)(n+2)/2} LG_1^{(3n-n^2)/2} = LG_2^{-1+2n+n^2}$, $(n=1,2,3\cdots)$. That is, corresponding to a series of transformations: $LG_1^2 LG_1^5 \to LG_2^7$, $LG_1^5 LG_1^9 \to LG_2^{14}$, $LG_1^9 LG_1^{14} \to LG_2^{23}$, and so on.

*Experiment.* — Next, the above analysis was verified by comparing our observations with their corresponding simulations. A straightforward way to verify the *full-field* transformation is spatial mode tomography employing spatial light modulator. However, it is difficult to exactly match beam parameters, such as beam size and divergence, required by radial mode projection in experiment. Therefore, for verifying the theory precisely and intuitively, the beam profiles of SHG fields from $z_0$ to $z_\infty$ plane were chosen as the observable characteristics. To record the beam profiles of generated SHG fields from $z_0$ to $z_\infty$ precisely, we employed a propagation tomography system based on a digital propagation technique [33] (see Ref. 32 for the details and MATLAB code). Further experimental details are provided in Appendix B.

Figure 1 shows the schematic of the experimental apparatus. A type-II SHG with two orthogonally polarized LG modes was used as the experimental platform. A narrow-linewidth 800 nm laser was first converted to a perfect horizontally polarized $TEM_{00}$ mode by passing it through a spatial filter in combination with a polarizing beam splitter (PBS). After this, the beam was incident on a phase-only spatial light modulator (SLM, Holoeye-VIS-080). Damman gratings based on complex amplitude modulation were used to generate a pair of LG modes simultaneously with the same $w_0$ but different spatial indices [32]. The two LG modes were converted to orthogonal polarizations via a half-wave plate (HWP) and then combined into a copropagating beam, or rather a vector mode, using a 4*f*-imaging system with a polarizing grating (PG) [44]. This vector mode was then loosely focused (*f* = 100 mm) into a 3-mm-long type-II PPKTP. Finally, the generated SHG field was sent to the propagation tomography system.

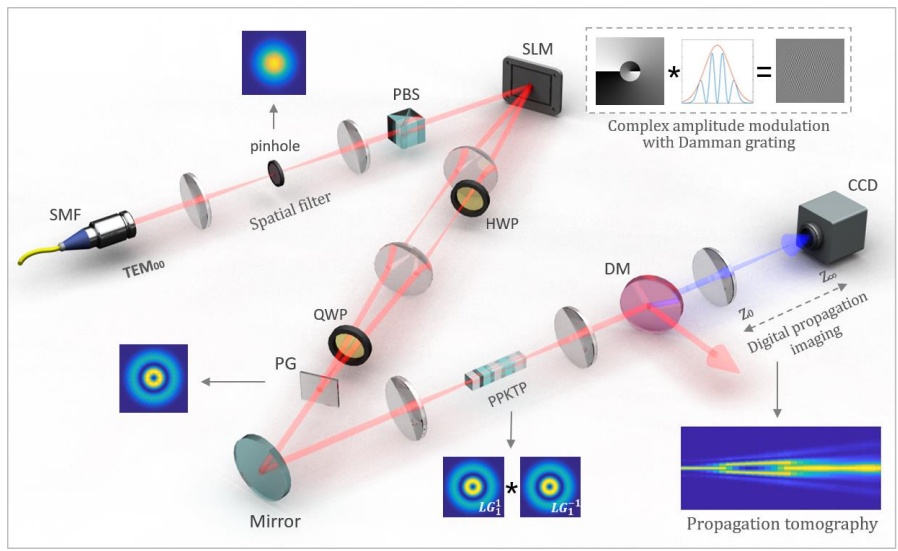

FIG. 2. Schematic of the experimental setup. The key components include the single mode fiber (SMF), polarizing beam splitter (PBS), spatial light modulator (SLM), half-wave plate (HWP), quarter-wave plate (QWP), polarizing grating (PG), and dichroic mirror (DM).

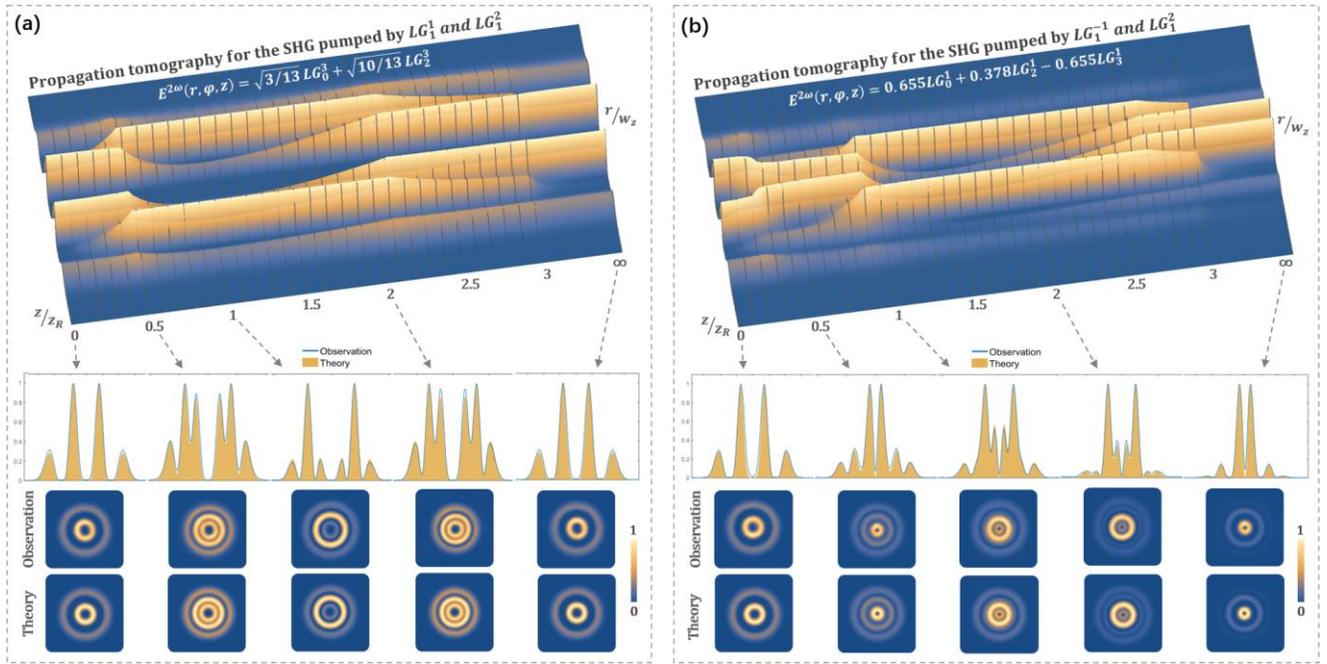

FIG. 3. Experimental results: (a) and (b) are propagation tomographies for the SHGs of $LG_1^1 LG_1^2$ and $LG_1^{-1} LG_1^2$, respectively. In the tomographies, the transverse unit and propagation distance are given in $r/w_z$ and $z/z_R$, respectively. The blue curves in the 3D intensity profiles and cross-sectional examples denote observed profiles. These are fitting data based on beam profiles observed upon propagation. The original data is provided in Appendix C.

The propagation tomographies of various SHG fields were recorded successfully with the help of the aforementioned digital propagation technique. Here, we focus on two representative examples: the SHGs of $LG_1^1 LG_1^2$ and $LG_1^{-1} LG_1^2$. These correspond to special cases (i) and (ii), respectively. Figures 3(a) and 3(b) show the observed propagation tomographies of the two examples (31 slices plotted using blue curves) and their corresponding theoretical predictions (yellow fillers), respectively. In both cases, the tomography results are in excellent agreement with the theoretical predictions. This confirms the accuracy of the theory. Moreover, the radial profiles of the two examples are the same at $z_0$ plane, but are different in the following propagation, i.e., one is self-imaging propagation while the other is not. These observations verify the theory discussed above. Further observations are provided in Appendix C.

*Discussion and Conclusions* — Based on the above demonstration, now it is clear that given an SFG pumped by two arbitrary LG modes, one can exactly simulate the spatial amplitude of the generated SHG field using Eq. (2) or (S3) for more general SFG. Then, the *full-field* selection rule in the form of Eq. (3) can be readily obtained via projective measurement onto LG modes. Except for the special case shown in Fig. 1(e), the generated SFG fields are usually not propagation-constant modes, but rather are superposition states that consist of a series of LG modes with the same $\ell$ but different $p$. On the basis of relations between LG modes and other spatial eigenmodes, one can further study the *full-field* transformation of other spatial modes, such as the mode and parity transformation of Ince Gauss modes in parametric processes. We also addressed the practical question of how to convert the frequency of a high-dimensional photonic state encoded with *full-field* spatial modes. According the data shown in Fig. 1(d), the key is to achieve a radially independent transformation, which can be excited by a flattop beam.

Our results in the case (i) indicate that the interesting phenomenon reported by Courtial *et al.* [3], i.e., observing SHG fields are self-imaging in propagation, applies not only to the SHG of a single pump, but also to the SHG of two pumps that obey Eq. (5). When $\ell_1$ and $\ell_2$ have the same sign, the odd-order complex coefficient $a_\eta$ becomes zero and thus the radial profile at the far field (except for overall enlarging) is the same as that at the generation plane. In the case (ii), the interesting number $N = 2p + |\ell|$ found in the special SHG of $LG_p^{+\ell} LG_p^{-\ell}$ can be explained by the fact that the far-field amplitude of an SFG field can be served as the auto- or cross-correlation function of two pump fields. This coincidence indicates SFG can be used to study spatial coherence between light beams with different wavelengths. From another perspective, according to Eqs. (2) and (3), the highest-order radial component of SHG dependents on both the azimuthal and radial indices of pump fields. Namely, the principle revealed in case (ii) explained the azimuthal-radial coupling in parametric interactions that has been found previously [27–30]. Besides, during the peer-review period of this work, we found there is another parallel work confirmed the principle of azimuthal-radial coupling in SHG [45].

In summary, the radial modal transition of LG modes in SFG was studied in detail both theoretically and experimentally. A general solution that describes the spatial complex amplitude of an SFG field pumped by two arbitrary LG modes was provided. Based on this solution, a *full-field* selection rule of LG modes for a given SFG can be obtained readily. Our general results effectively explain interesting phenomena and extend the cognition reported in previous relevant works. More important, the theory, on the one hand, provides a basis for quantum and nonlinear optical research involving parametric upconversion of complex structured light; and on the other hand, paves the way for future work on full-field transformation of spatial modes in other nonlinear interactions.

**ACKNOWLEDGMENT**

We thank Prof. Cheng-Wei Qiu from National University of Singapore for fruitful discussions. We thank professional and useful inputs from anonymity referees. This work was supported by the National Natural Science Foundation of China (Grant Nos. 11934013, 61975047, and 11874102).

# Appendix A  Detailed Theoretical Framework

## 1. Full spatial wavefunction of a sum frequency generation (SFG) field pumped by Laguerre-Gaussian (LG) modes

The wavefunction of the LG mode used in simulations is given in cylindrical coordinates $\{r,\varphi,z\}$ by

$$LG_p^\ell(r,\varphi,z) = \sqrt{\frac{2p!}{\pi(p+|\ell|)!}}\frac{1}{w(z)}\left(\frac{\sqrt{2}r}{w(z)}\right)^{|\ell|}\exp\left(\frac{-r^2}{w_z^2}\right)\times L_p^{|\ell|}\left(\frac{2r^2}{w_z^2}\right)\exp\left[-i\left(kz+\frac{kr^2}{2R_z}+\ell\varphi-i\phi_g\right)\right], \quad (S1)$$

where $w_z = w_0\sqrt{1+(z/z_R)^2}$, $R_z = z^2 + z_R^2/z$, and $\phi_g = (2p+|\ell|+1)\arctan(z/z_R)$ denote the beam waist, radius of curvature, and Gouy phase upon propagation (here $z_R = kw_0^2/2$ is the Rayleigh length), respectively, and $L_p^{|\ell|}(\cdot)$ is the Laguerre polynomial with mode orders $p$ and $|\ell|$, given by $L_p^{|\ell|}(\gamma) = \sum_{k=0}^{p}\left[(|\ell|+p)!(-\gamma)^k\right]/\left[(|\ell|+k)!k!(p-k)!\right]$. In the main text, Eq. (S1) is abbreviated as $LG_p^\ell(r,\varphi,z) = g(r,z;k,w_0,p,\ell)L_p^{|\ell|}(\gamma)\exp(-i\ell\varphi)$, which corresponds to Eq. (1).

For an SFG driven by two LG modes $LG_{p_1}^{\ell_1}(r_0,\varphi_0)$ with $k_1$ and $LG_{p_2}^{\ell_2}(r_0,\varphi_0)$ with $k_2$, the wave source of the SFG field (i.e., nonlinear polarization) is excited by the quadratic beating field of the two LG modes with $k_3 = k_1 + k_2$, given by

$$E_{SFG}(r_0,\varphi_0) = LG_{p_1}^{\ell_1}(r_0,\varphi_0)LG_{p_2}^{\ell_2}(r_0,\varphi_0)$$
$$= \frac{2}{\pi}\sqrt{\frac{p_1!p_2!}{(|\ell_1|+p_1)!(|\ell_2|+p_2)!}}\frac{(\sqrt{2}r)^{|\ell_1|+|\ell_2|}}{w_1^{|\ell_1|+|\ell_2|}w_2^{|\ell_1|+|\ell_2|}}\exp\left[-r\left(\frac{1}{w_1^2}+\frac{1}{w_2^2}\right)\right]\exp\left[-i(\ell_1+\ell_2)\right]L_{p_1}^{|\ell_1|}\left(\frac{2r^2}{w_z^2}\right)L_{p_2}^{|\ell_2|}\left(\frac{2r^2}{w_z^2}\right), \quad (S2)$$

where $w_1$ and $w_2$ denote the beam waists of two pumps. Therefore, the traveling wave equation of the SFG field $E_{SFG}(r,\varphi,z)$ can be derived by using the Collins propagator with $E_{SFG}(r_0,\varphi_0)$ as the pupil function. This produces

$$E_{SFG}(r,\varphi,z) = \frac{i}{\lambda z}\exp(-ik_3 z)\int r_0 dr_0 \int d\varphi_0 E^{2\omega}(r_0,\varphi_0)\exp\left\{-\frac{ik_3}{2z}[r_0^2 - 2rr_0\cos(\varphi-\varphi_0)+r^2]\right\}$$
$$= \frac{1}{\lambda z}\sqrt{\frac{2^{|\ell_1|+|\ell_2|+4}i^{2|\ell_1+\ell_2|+2}p_1!p_2!}{(|\ell_1|+p_1)!(|\ell_2|+p_2)!}}\frac{1}{w_1^{|\ell_1|+1}w_2^{|\ell_2|+1}}\frac{\alpha^{|\ell_1+\ell_2|}}{2\beta^{|\ell_1|+|\ell_2|+|\ell_1+\ell_2|+2}}\exp\left(-\frac{\alpha^2}{4\beta^2}\right)$$
$$\times \exp\left\{-i\left[\frac{k_3 r^2}{2z}+k_3 z+(\ell_1+\ell_2)\varphi\right]\right\}\sum_{j=0}^{p_1+p_2}c_j\frac{1}{\beta^{2j}}\left(\frac{|\ell_1|+|\ell_2|-|\ell_1+\ell_2|}{2}+j\right)!L_{\frac{|\ell_1|+|\ell_2|-|\ell_1+\ell_2|}{2}+j}^{|\ell_1+\ell_2|}\left(\frac{\alpha^2}{4\beta^2}\right) \quad (S3)$$

where $\alpha = kr/z$, $\beta = \sqrt{(1/w_1^2 + 1/w_2^2)+ik/2z}$, and $c_j$ denotes the coefficient of the series $L_{p_1}^{|\ell_1|}(\cdot)L_{p_2}^{|\ell_2|}(\cdot)$. If we assume that $w_1 = w_2 = w_0$, by substituting $m = \ell_1 + \ell_2$ and $n = (|\ell_1|+|\ell_2|-|\ell_1+\ell_2|)/2$ into Eq. (S3), we can reformulate the equation as

$$E_{SFG}(r,\varphi,z) = \frac{1}{\lambda z}\sqrt{\frac{2^{|\ell_1|+|\ell_2|+4}i^{2|\ell_1+\ell_2|+2}p_1!p_2!}{(|\ell_1|+p_1)!(|\ell_2|+p_2)!}}\frac{1}{w_0^{|\ell_1|+|\ell_2|+2}}\frac{\alpha^{|\ell_1+\ell_2|}}{2\beta^{|\ell_1|+|\ell_2|+|\ell_1+\ell_2|+2}}\exp\left(-\frac{\alpha^2}{4\beta^2}\right)\exp\left[-i\left(\frac{k_3 r^2}{2z}+k_3 z+m\varphi\right)\right]L_{SFG}(r,z)$$

$$L_{SFG}(r,z) = \sum_{j=0}^{p_1+p_2}c_j\frac{1}{\beta^{2j}}(n+j)!L_{n+j}^{|m|}\left(\frac{\alpha^2}{4\beta^2}\right)\xrightarrow[\zeta_z = \frac{\alpha^2}{4\beta^2}]{q_j(z)=\frac{1}{\beta^{2j}}(n+j)!}\sum_{j=0}^{p_1+p_2}q_j(z)L_{n+j}^{|m|}(\zeta_z) \quad (S4)$$

For the particular case of a type-II SHG, Eq. (S4) can be further simplified by assuming that $k_1 = k_2$. The result is abbreviated as $E^{2\omega}(r,\varphi,z) = g^{2\omega}(r,z;k_3,w_0,p_{1,2},\ell_{1,2})L^{2\omega}(r,z)\exp(-im\varphi)$ in the main text and corresponds to Eq. (2). By projecting Eq. (S4) onto LG modes with the same azimuthal index $m$ but different radial index $\eta$, we can obtain the full-field selection rule with respect to LG mode, given by $E^{2\omega}(r,\varphi,z) = \sum_{n=0}^{n+j}a_n LG_n^m(r,\varphi,z)$, i.e., Eq. (3) in the main text, where $a_n$ denotes complex coefficients that describe the weighting factor and intramodal phase of the superposition mode. For instance, if $\ell_1 \geq 0$, $\ell_2 \geq 0$ and $p_1 = p_2 = 1$, the first three complex coefficients $a_n$ ($\eta = 0,1,2$) are given by

$$a_0 = \sqrt{2/\pi}\left\{\ell_1^2 w_1^4 + [1+\ell_2]w_2^2\left[-2w_1^2+\ell_2 w_2^2\right]+\ell_1\left[w_1^4 - 2(1+\ell_2)w_1^2 w_2^2\right]\right\}\Gamma(1+\ell_1+\ell_2)/(w_1^2+w_2^2)^2\sqrt{(\ell_1+\ell_2)!};$$
$$a_1 = \sqrt{2/\pi}(w_1 - w_2)(w_1 + w_2)\left[(1+\ell_1)w_1^2 - (1+\ell_2)w_2^2\right]\sqrt{\Gamma(2+\ell_1+\ell_2)}/(w_1^2+w_2^2)^2;$$
$$a_2 = 2w_1^2 w_2^2 \sqrt{\Gamma(3+\ell_1+\ell_2)}/\sqrt{\pi}(w_1^2+w_2^2)^2. \quad (S5)$$

## 2. Far-field amplitudes of SFG fields

At the far field, i.e., $z \to z_\infty$, Eq. (S4) becomes the Fourier transform of the SFG source $E_{SFG}(r_0, \varphi_0)$ and $\zeta_z \to \mathrm{R}$, which is given by

$$\mathcal{F}[E_{SFG}(r_0,\varphi_0)] = \int_0^{2\pi}\int_0^\infty E_{SFG}(r_0,\varphi_0)\exp\left[-i2\pi r_0 r \cos(\varphi-\varphi_0)\right]r_0 dr_0 d\varphi_0$$

$$= \sqrt{\frac{2^{|\ell_1|+|\ell_2|+4}i^{2|m|}p_1!p_2!}{(|\ell_1|+p_1)!(|\ell_2|+p_2)!}} \frac{1}{w_0^{|\ell_1|+|\ell_2|+2}} \frac{\alpha^{|m|}}{2\beta^{|\ell_1|+|\ell_2|+|m|+2}}\exp\left(-\frac{\alpha^2}{4\beta^2}\right)\exp(-im\varphi)L_{SHG}(r,z_\infty)$$

$$L_{SFG}(r,z_\infty) = \sum_{j=0}^{p_1+p_2} q_j L_{n+j}^{|m|}(\zeta) = \sum_{j=0}^{p1+p2}\left(c_j \frac{1}{\beta^{2j}}(n+j)!L_{n+j}^{|m|}(\zeta)\right), \tag{S6}$$

where $\zeta$ denotes a real function ($\zeta_z \to \mathrm{R}$). If we further assume that $p_1 = p_2$, the radial governing term $L_{SFG}(r,z_\infty)$ can be further factorized as

$$L_{SFG}(r,z_\infty) \xrightarrow{p_1=p_2} \begin{cases} L_p^{|\ell_1|}(\zeta)L_p^{|\ell_2|}(\zeta), & (\text{for } \ell_1 \times \ell_2 \geq 0) \\ \frac{(p+n)!}{p!}L_{p+n}^{|m|}(\zeta)L_p^0(\zeta), & (\text{for } \ell_1 \times \ell_2 < 0) \end{cases}. \tag{S7}$$

Here, the arrow symbol denotes substitution calculation. For the particular case of a type-II SHG, Eqs. (S6) and (S7) are abbreviated in the main text as Eqs. (4) and (5), respectively.

## Appendix B   Experimental Details

### 1. Generation of two arbitrary LG modes via phase-only spatial light modulation (SLM)

The most common way to generate LG modes is using phase-only holography, where two spatial indices of a target LG mode need to be coded into the illumination light. To shape the azimuthal mode, a phase-only make carrying a twisted phase is competent. In contrast, shaping the radial mode via phase-only holography as performed in this experiment was more difficult. This task generally requires complex amplitude modulation.

The principle of complex amplitude modulation is based on the dependence of diffraction efficiency on the phase depth of a blazed grating (further details of the principle and corresponding MATLAB code are provided in Ref. 31). Figure S1(a) shows the measured dependence of the SLM (Holoeye VIS-080) used in the experiment. Based on this dependence, one could add the intensity mask of a target LG mode to the phase-only hologram. If a plane wave were used as the illumination light, the principle shown above would be enough. However, the illumination light usually offers the TEM$_{00}$ mode shown in Fig. S1(b). Therefore, the intensity mask was corrected based on the observed Gaussian envelope with $w_0 = 1.403$ mm. Figure S1(c) compares the beam profiles of the generated LG modes before and after the illumination correction and corresponding holograms are shown in Fig. S1(d). After the correction, the generated LG mode was in excellent agreement with the theoretical expectation of the target mode. Finally, one must carefully select an optimal beam waist ($w_0^{LG}$), where the principle should be to pursue a high generation efficiency without exceeding the envelope of illumination, as shown in Figs. S1(e) and S1(f). For this, we defined a fill factor $n$, given by $w_0^{LG} = 1.403n$. In addition, the Rayleigh lengths of generated LG modes are defined experimentally based on their $w_0^{LG}$.

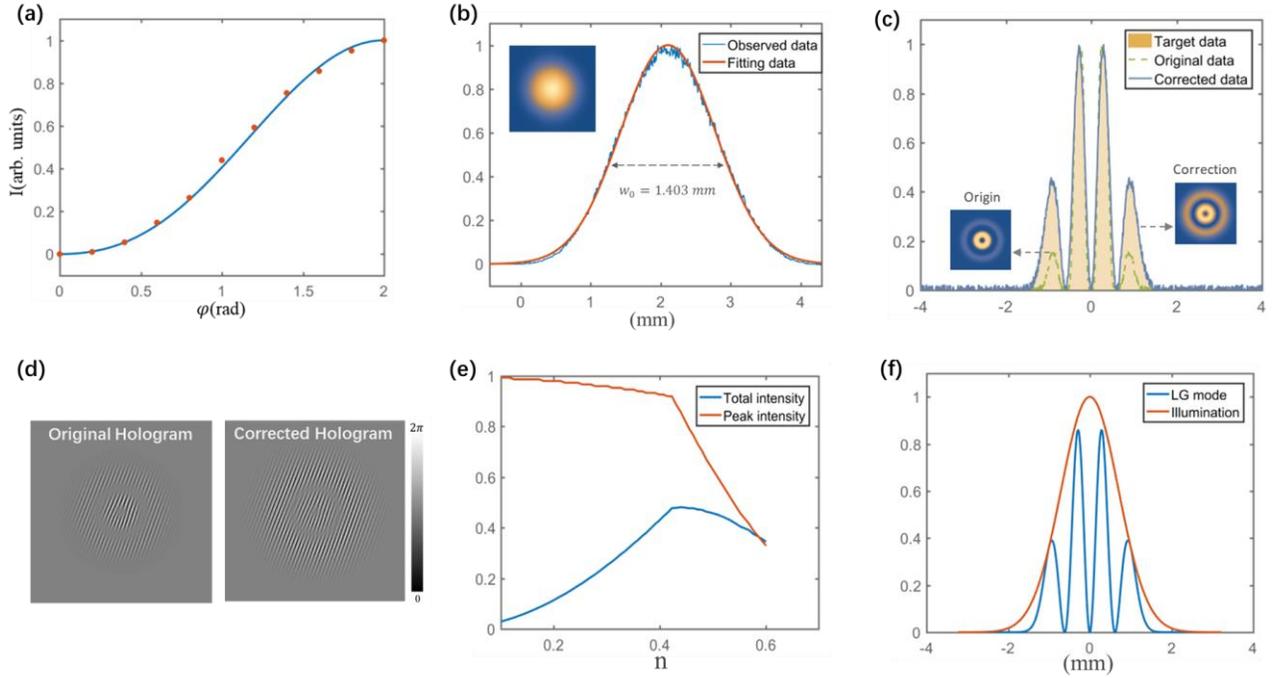

FIG. S1. Experimental details for generation of exact LG modes via phase-only SLM: (a) The dependence of the 1st order diffraction efficiency on the phase depth of the blazed grating. The blue curve denotes the theory and the red points are observations. (b) Measured beam waist of the Gaussian illumination light. (c) The beam profiles of the generated LG mode before and after the illumination correction. (d) The LG mode generation phase masks before and after the illumination correction. (e) and (f) The dependence of the generation efficiency on the beam waist of the target LG mode.

## 2. Digital propagation imaging system for propagation tomography

Figure S2 shows the digital propagation imaging system experimental apparatus. The beam size of the SHG field generation plane was first doubled using a 4f imaging lens set (focal lengths: 100 mm and 200 mm). Then, the enlarged generation plane was relayed to the CCD surface by another 4f imaging system (two 300 mm lenses). To achieve digital propagation, a SLM for a 400 nm laser (Holoeye UV-099) was mounted on the Fourier plane of the relay imaging system. Further details of the principle of digital propagation and corresponding MATLAB code are provided in Ref. 31.

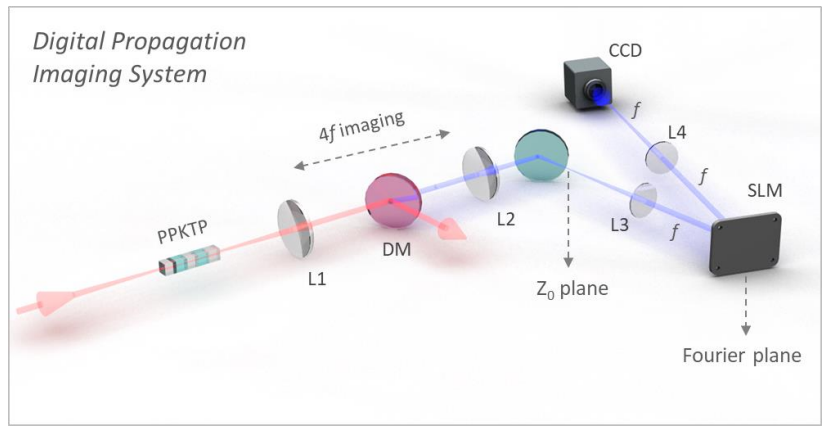

FIG. S2. Schematic of the digital propagation imaging system. The key components include the lenses (L1–L4), spatial light modulator (SLM) and dichroic mirror (DM). Here, the focal lengths of L1, L2, L3, and L4 are 100 mm, 200 mm, 300 mm, and 300 mm, respectively.

# Appendix C    Original data of propagation tomography

Here we provide additional results recorded by propagation tomography, i.e., the radial structures of observed SHG fields from $z_R = 0$ to $z_R = 3$ (left column), and their corresponding theoretical simulations (right column). The resolution of the tomography is 0.1 $z_R$, i.e., 30 propagation slices from $z = 0$ to $z = 3z_R$.

Data for the case (i)

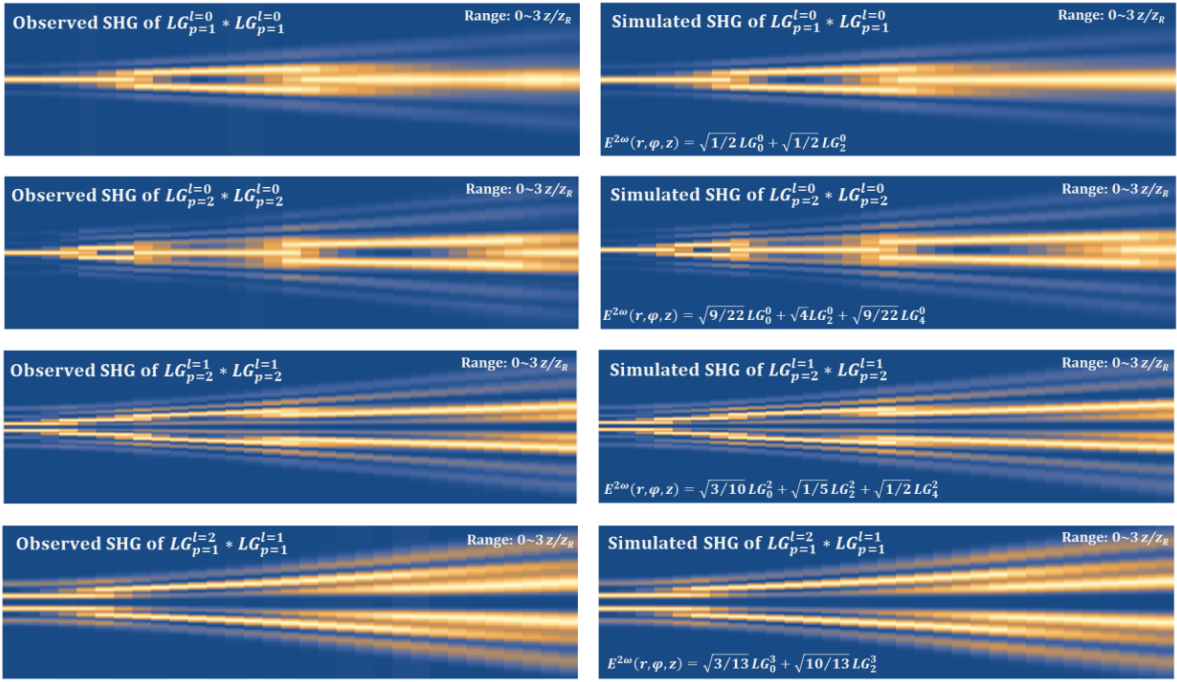

Data for the case (ii)

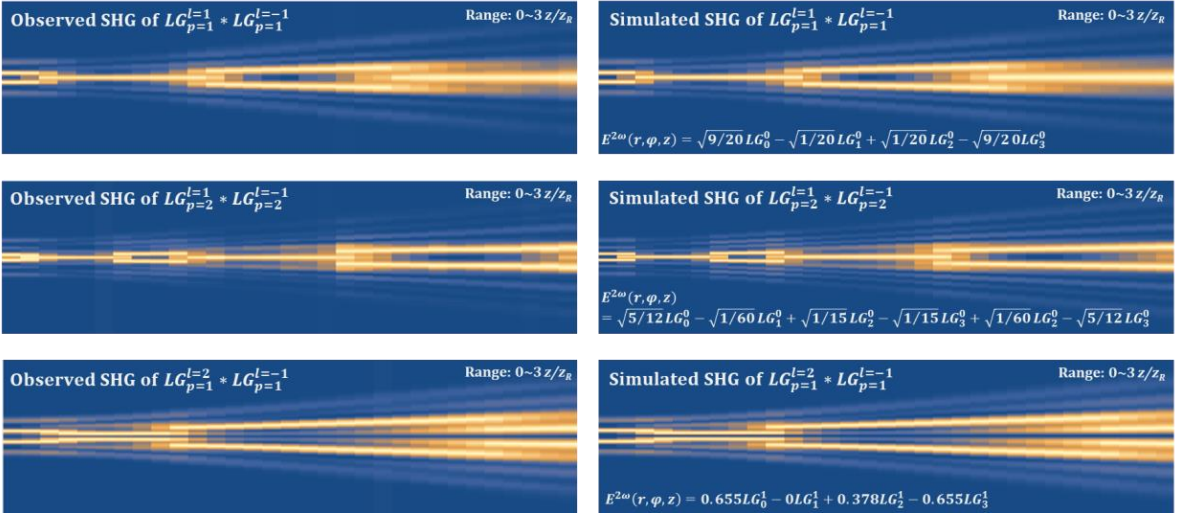

FIG. S3. Additional results recorded by propagation tomography.